\documentclass[conference]{IEEEtran}
\usepackage{graphicx} % Required for inserting images
\usepackage[percent]{overpic}
\usepackage{cite}
\usepackage[colorlinks=true, linkcolor=black, citecolor=black, urlcolor=blue, bookmarks=false]{hyperref}
\usepackage{amsmath, amsfonts, amssymb}
\usepackage{algorithm}
\usepackage{algpseudocode}
\usepackage{xcolor}

\IEEEoverridecommandlockouts
\IEEEaftertitletext{\vspace{-1\baselineskip}}

\title{Accurate and Fast Channel Estimation for Fluid Antenna Systems with Diffusion Models}

\author{
    \IEEEauthorblockN{
        Erqiang Tang\IEEEauthorrefmark{1},
        Wei Guo\IEEEauthorrefmark{1},
        Hengtao He\IEEEauthorrefmark{2},
        Shenghui Song\IEEEauthorrefmark{1}, \\
        Jun Zhang\IEEEauthorrefmark{1}, \textit{Fellow, IEEE},
        Khaled B. Letaief\IEEEauthorrefmark{1}, \textit{Fellow, IEEE}
    }
    \IEEEauthorblockA{
        \IEEEauthorrefmark{1}Dept. of Electronic and Computer Engineering, The Hong Kong University of Science and Technology, Hong Kong \\
        \IEEEauthorrefmark{2}National Mobile Communications Research Laboratory, Southeast University, Nanjing, China \\
        Emails: etangaa@connect.ust.hk, eeweiguo@ust.hk, hehengtao@seu.edu.cn, \{eeshsong, eejzhang, eekhaled\}@ust.hk
    }

}

\begin{document}

\maketitle

\begin{abstract}
Fluid antenna systems (FAS) offer enhanced spatial diversity for next-generation wireless systems. However, acquiring accurate channel state information (CSI) remains challenging due to the large number of reconfigurable ports and the limited availability of radio-frequency (RF) chains---particularly in high-dimensional FAS scenarios. To address this challenge, we propose an efficient posterior sampling-based channel estimator that leverages a diffusion model (DM) with a simplified U-Net architecture to capture the spatial correlation structure of two-dimensional FAS channels. The DM is initially trained offline in an unsupervised way and then applied online as a learned implicit prior to reconstruct CSI from partial observations via posterior sampling through a denoising diffusion restoration model (DDRM). To accelerate the online inference, we introduce a skipped sampling strategy that updates only a subset of latent variables during the sampling process, thereby reducing the computational cost with minimal accuracy degradation. Simulation results demonstrate that the proposed approach achieves significantly higher estimation accuracy and over 20× speedup compared to state-of-the-art compressed sensing-based methods, highlighting its potential for practical deployment in high-dimensional FAS.
\end{abstract}


\section{Introduction}

The fluid antenna system (FAS) has recently been proposed as an emerging antenna technology for next-generation wireless communication systems~\cite{wong2020fluid}. Broadly, it refers to any form of movable or non-movable antenna systems with position reconfiguration capability~\cite{zhu2024historical}. Unlike conventional multiple-input-multiple-output (MIMO) systems with antennas fixed in a predefined array, FAS can dynamically adjust their antennas to the most favorable locations within a designated space. Leveraging these additional spatial degrees of freedom, FAS has demonstrated significant performance gains in both point-to-point and multi-user communication scenarios~\cite{new2023fluid, zhu2023movable}.

These promising benefits of FAS rely heavily on the optimal positioning of its fluid antennas (FAs), which in turn requires accurate channel state information (CSI) at all available ports \cite{zhu2023movable, liao2025joint}. However, channel estimation for FAS is challenging. On one hand, FAS ports are densely deployed to achieve high spatial resolution, resulting in an extremely high-dimensional channel vector to estimate. On the other hand, only a few ports can be accessed at a time due to limited radio frequency (RF)-chains available. Therefore, huge switching overhead would be incurred if each port is estimated individually.

To address these challenges, it is essential to obtain the full CSI from only partial port observations. Existing studies can be broadly divided into two main categories. The first leverages the inherent sparsity of wireless channels, with its design principles rooted in traditional MIMO systems, and has motivated the development of numerous compressed sensing (CS)-based methods \cite{xiao2024channel, xu2024sparse}. Nevertheless, these methods significantly depend on the channel sparsity assumption, which only holds in channel environments with limited scatterers. Furthermore, the iterative nature of most CS-based algorithms leads to considerable computational complexity, particularly when the dimensionality of the FAS channel grows. The other category of approaches leverages the strong spatial correlation among FAS ports due to their dense configurations. For instance, \cite{zhang2024successive} treats one-dimensional (1D) FAS channels as Gaussian random processes and proposes a kernel-based sampling and regression technique to successively reconstruct full CSI from limited port samples. However, generalizing this approach to high-dimensional FAS channels is non-trivial since it is difficult to construct a high-dimensional random process and perform channel estimation accordingly.

These limitations motivate the use of learning-based methods to directly capture the inherent port correlation structure of FAS channels from data, which bypass the need for restrictive assumptions on channel sparsity or explicit modeling of complex correlations in higher-dimensional settings. In particular, diffusion models (DMs) \cite{ho2020denoising} have shown exceptional ability in capturing complex data structures and generating high-quality synthetic samples. Motivated by their recent success in various image restoration \cite{jalal2021robust, kawar2022denoising} and wireless communication problems \cite{yu2024bayes, arvinte2022mimo}, we propose an efficient DM-aided posterior sampling approach for accurate two-dimensional (2D) FAS channel estimation.

Specifically, we design a lightweight DM based on a simplified U-Net architecture to capture the spatial correlations of 2D FAS channels, inspired by their structural similarity to natural images. The DM is trained offline in an unsupervised manner using generated channel samples. By treating the trained DM as a learned implicit prior, we apply it online during channel estimation to reconstruct FAS channels from partial observations via posterior sampling under the denoising diffusion restoration model (DDRM) framework \cite{kawar2022denoising}. Furthermore, to speed up the sampling process, we propose to use an acceleration scheme by skipped sampling, which only samples partial latent variables during the reverse process. This enables real-time channel estimation without compromising accuracy in practical scenarios.

% \vspace{-1em}

\section{System Model and Preliminaries}

In this section, we first present the considered 2D FAS channel model and formulate the corresponding channel estimation problem. Then, we provide necessary preliminaries on diffusion models, which serve as the foundation for our proposed method.

% \vspace{-1.2em}

\subsection{System Model}

\begin{figure}[!t]
    \centering
    \includegraphics[width=0.92\linewidth]{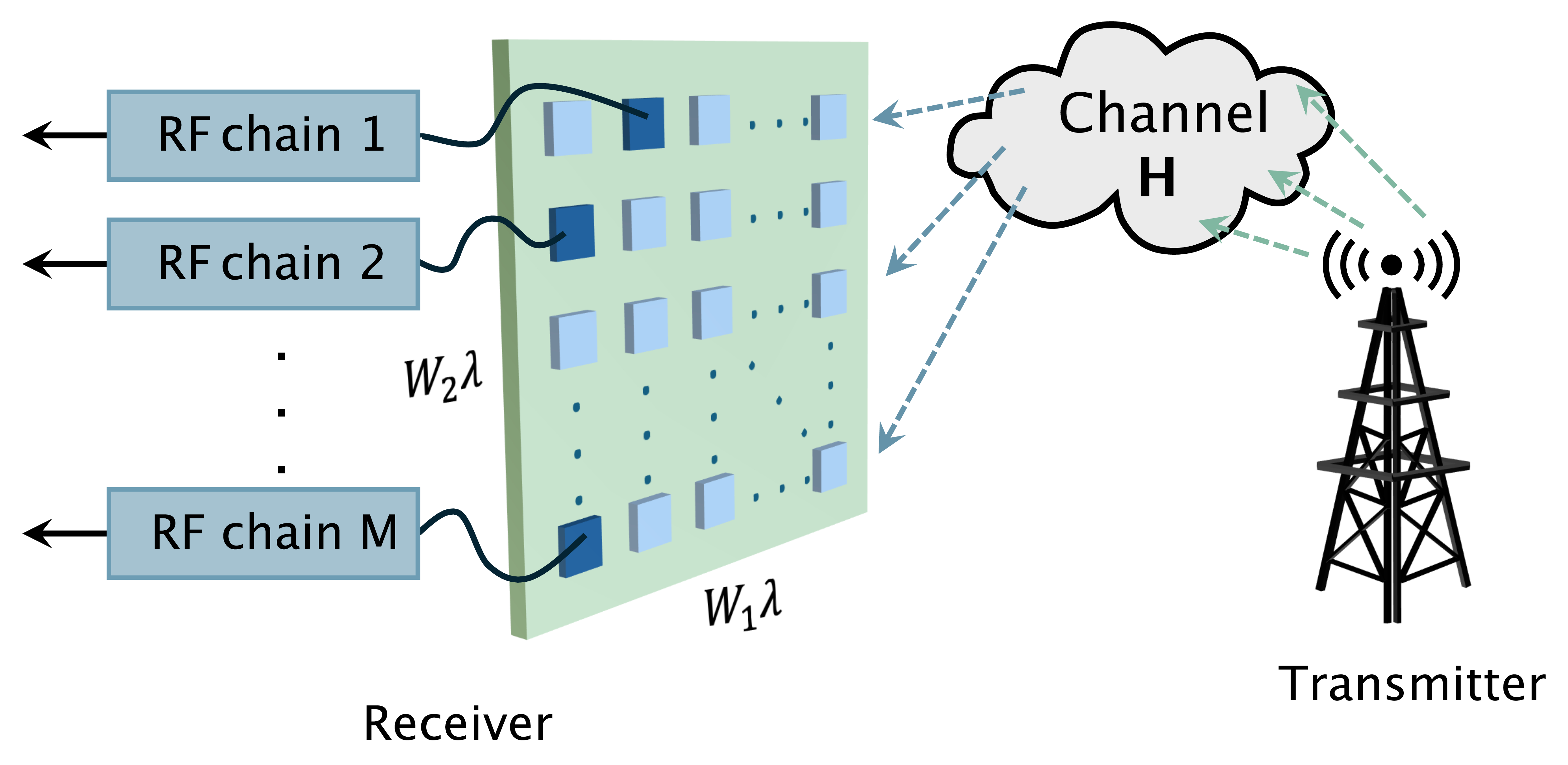}
    \caption{An illustration of the point-to-point system considered, where the receiver is equipped with an $N$-port 2D FAS and has $M$ RF-chains.}
    \label{fig:system_model}
\end{figure}

As shown in Fig.\,\ref{fig:system_model}, we consider a narrowband point-to-point communication system, where the transmitter is equipped with a single fixed-position antenna while the receiver employs a 2D reconfigurable pixel-based FAS. The FAS surface spans a rectangular region of size $W = W_1 \lambda \times W_2 \lambda$, where $\lambda$ denotes the carrier wavelength. The surface is uniformly discretized into $N = N_1 \times N_2$ ports among which the FAs can switch. Additionally, the FAS consists of $M$ RF-chains ($M \ll N$), each connected to a dedicated FA that can be repositioned to any of the $N$ available ports for signal reception.

Let $\mathbf{H} \in \mathbb{C}^{N_1 \times N_2}$ denote the complex channel responses across all ports. Assuming the propagation environment contains $N_p$ dominant signal paths due to finite scatterers, the channel matrix $\mathbf{H}$ can be modeled as
\begin{equation}
    \mathbf{H} = \sqrt{\frac{1}{N_p}} \sum_{i=1}^{N_p} g_i \mathbf{a}_x (\theta_i, \phi_i) \mathbf{a}_y(\theta_i,\phi_i)^T,
    \label{eq:chan_model}
\end{equation}
where $g_i$ represents the complex gain of each path, and $\theta_i$ and $\phi_i$ denote the corresponding azimuth and elevation angles-of-arrival (AoAs). The steering vectors $\mathbf{a}_x (\theta, \phi)$ and $\mathbf{a}_y (\theta, \phi)$ are respectively defined as\footnote{In fact, $\mathbf{a}_y$ is independent of the azimuth angle $\theta$, but we still keep $\theta$ in its argument for notation consistency.}
\begin{equation}
\begin{aligned}
\mathbf{a}_x(\theta,\phi) &= \left[1, e^{-j 2\pi \frac{W_1}{N_1-1} \cos\phi\sin\theta}, \cdots, e^{-j 2\pi W_1 \cos\phi\sin\theta} \right]^T, \\
\mathbf{a}_y(\theta,\phi) &= \left[1, e^{-j 2\pi \frac{W_2}{N_2-1} \sin\phi}, \cdots, e^{-j 2\pi W_2 \sin\phi} \right]^T,
\end{aligned}
\end{equation}

Assume that during the estimation period, $L$ time slots are allocated for pilot transmission. At each time slot $l$, the receiver FAS chooses $M$ out of $N$ ports for observation, which can be characterized by a binary selection matrix $\mathbf{S}_l \in \{0,1\}^{M\times N}$, also known as the switch matrix \cite{zhang2024successive}. Each row of $\mathbf{S}_l$ is picked from the identity matrix $\mathbf{I}_N$, such that if the $m$-th FA is switched to the $n$-th port for pilot reception during time slot $l$, then the $m$-th row of $\mathbf{S}_l$ corresponds to the $n$-th row of $\mathbf{I}_N$. As a result, the received signal at time slot $l$, denoted by $\mathbf{y}_l \in \mathbb{C}^{M\times 1}$, can be expressed as
\begin{equation}
    \mathbf{y}_l = \mathbf{S}_l \tilde{\mathbf{h}} x_l + \mathbf{n}_l,
\end{equation}
where $\tilde{\mathbf{h}} = \mathrm{vec}(\mathbf{H})$ is the flattened channel, $x_l$ represents the transmitted pilot symbol, and $\mathbf{n}_l \sim \mathcal{CN}(\mathbf{0},\sigma_n^2 \mathbf{I}_M)$ is the additive white Gaussian noise (AWGN). Without loss of generality, we set $x_l = 1$ for all $l \in \{1,\ldots,L\}$. Aggregating the received signals over the all time slots yields
\begin{equation}
    \tilde{\mathbf{y}} = \widetilde{\mathbf{S}} \, \tilde{\mathbf{h}} + \tilde{\mathbf{n}},
    \label{eq:rx_signal_complex}
\end{equation}
where $\tilde{\mathbf{y}} = [\mathbf{y}_1^T,\ldots,\mathbf{y}_L^T]^T \in \mathbb{C}^{LM \times 1}$ is the stacked received signal across all $L$ time slots, $\tilde{\mathbf{S}} = [\mathbf{S}_1^T,\ldots,\mathbf{S}_L^T]^T \in \mathbb{C}^{LM\times N}$ is the overall port selection matrix, and $\tilde{\mathbf{n}} = [\mathbf{n}_1^T,\ldots,\mathbf{n}_L^T]^T \in \mathbb{C}^{LM\times 1}$ is the stacked noise vector.

To facilitate learning-based channel estimation, we consider the equivalent real-valued form of \eqref{eq:rx_signal_complex}. Specifically, we define $\mathbf{y} = [\Re(\tilde{\mathbf{y}})^T, \Im(\tilde{\mathbf{y}})^T]^T \in \mathbb{R}^{\bar{M} \times 1}$, $\mathbf{h} = [\Re(\tilde{\mathbf{h}})^T, \Im(\tilde{\mathbf{h}})^T]^T \in \mathbb{R}^{\bar{N} \times 1}$, $\mathbf{n} = [\Re(\tilde{\mathbf{n}})^T, \Im(\tilde{\mathbf{n}})^T]^T \in \mathbb{R}^{\bar{M} \times 1}$, and
\begin{equation}
    \mathbf{S} = \begin{bmatrix}
    \Re(\widetilde{\mathbf{S}}) & -\Im(\widetilde{\mathbf{S}}) \\
    \Im(\widetilde{\mathbf{S}}) & \Re(\widetilde{\mathbf{S}})
    \end{bmatrix} \in \mathbb{R}^{\bar{M} \times \bar{N}},
\end{equation}
with $\bar{M} = 2LM$ and $\bar{N} = 2N$. Under these notations, the equivalent real-valued received signal model is given by
\begin{equation}
    \mathbf{y} = \mathbf{S} \mathbf{h} + \mathbf{n}.
    \label{eq:rx_signal_real}
\end{equation}
The goal of channel estimation is to recover the full CSI $\mathbf{h}$ from the noisy partial port observations $\mathbf{y}$ given $\mathbf{S}$.

This problem can be effectively tackled from a Bayesian perspective. Given noisy observations $\mathbf{y}$, the posterior distribution of $\mathbf{h}$ can be formulated as $p(\mathbf{h}|\mathbf{y}) \propto p(\mathbf{y}|\mathbf{h})p(\mathbf{h})$, where the likelihood $p(\mathbf{y}|\mathbf{h})$ is defined by the signal model~\eqref{eq:rx_signal_real}, while the prior $p(\mathbf{h})$ is intended to encapsulate the known characteristics of the FAS channel. From this posterior, point estimates such as the minimum mean squared error (MMSE) estimate $\hat{\mathbf{h}}^{\text{MMSE}} = \mathbb{E}[\mathbf{h}|\mathbf{y}]$ can then be derived.

An appropriate prior is crucial for accurate channel estimation since the inverse problem in~\eqref{eq:rx_signal_real} is under-determined. Existing works typically use hand-crafted priors such as sparsity in a transformed domain \cite{xu2024sparse}. However, these simplified priors often fail to represent more complex correlation structures inherent in FAS channels. Alternatively, generative models like DMs can be utilized as a learned implicit prior, which are capable of capturing the spatial structures of FAS channels in a purely data-driven manner.

\subsection{Preliminaries on Diffusion Models}

Given observed samples $\mathbf{h}$ drawn from an unknown distribution $q(\mathbf{h})$, generative modeling seeks to learn a parametrized model $p_{\boldsymbol{\theta}}(\mathbf{h})$ that approximates the true data distribution. DMs \cite{ho2020denoising}, in particular, are a class of generative models with a Markov chain structure consisting of $T$ steps:
\begin{equation}
    p_{\boldsymbol{\theta}}(\mathbf{h}_{0:T}) = p(\mathbf{h}_T) \prod_{t=1}^T  p_{\boldsymbol{\theta}}(\mathbf{h}_{t-1}|\mathbf{h}_t),
\end{equation}
where $\mathbf{h}_{1:T}$ are latent variables of the same dimensionality as the original data $\mathbf{h}_0$ and $t$ indexes the hierarchical level of the latent variables. The learned distribution $p_{\boldsymbol{\theta}}(\mathbf{h})$ can be obtained by marginalizing over all latent variables $p_{\boldsymbol{\theta}}(\mathbf{h}_0) = \int p_{\boldsymbol{\theta}}(\mathbf{h}_{0:T}) d \mathbf{h}_{1:T}$.

This generative process aims to reverse a fixed forward diffusion process. Specifically, starting from a data sample $\mathbf{h}_0 \sim q(\mathbf{h}_0)$, the DM's forward process is defined as
\begin{equation}
    q(\mathbf{h}_t|\mathbf{h}_{t-1}) = \mathcal{N}(\mathbf{h}_t; \sqrt{1-\beta_t} \mathbf{h}_{t-1}, \beta_t \mathbf{I}), \: t = 1,\ldots, T,
\end{equation}
where $\beta_1,\ldots,\beta_T$ represent a predetermined noise schedule. A key property of this forward process is that sampling $\mathbf{h}_t$ directly from $\mathbf{h}_0$ at any $t$ is feasible via
\begin{equation}
    \mathbf{h}_t = \sqrt{\bar{\alpha}_t} \mathbf{h}_0 + \sqrt{1- \bar{\alpha}_t} \boldsymbol{\epsilon}_t, \: \boldsymbol{\epsilon}_t \sim \mathcal{N}(0,\mathbf{I}),
    \label{eq:ddpm_diffusion}
\end{equation}
where we let $\alpha_t = 1- \beta_t$ and $\bar{\alpha}_t = \prod_{i=1}^t \alpha_i$. The noise schedule $\{\beta_t\}_{t=1}^T$ is carefully designed such that $\bar{\alpha}_t \approx 0$, ensuring that $\mathbf{h}_T$ eventually approaches an isotropic Gaussian distribution $\mathcal{N}(0,\mathbf{I})$. Intuitively, the forward process progressively corrupts clean data samples into pure noise, while the reverse process aims to denoise from this noise to generate data samples. According to \cite{ho2020denoising}, both processes share the same functional form when $\beta_t$ is small, allowing us to parametrize $p_{\boldsymbol{\theta}}(\mathbf{h}_{t-1}|\mathbf{h}_t) = \mathcal{N}(\mathbf{h}_{t-1}; \boldsymbol{\mu}_{\boldsymbol{\theta}}(\mathbf{h}_t,t), \boldsymbol{\Sigma}_{\boldsymbol{\theta}}(\mathbf{h}_t,t))$.

To train a diffusion model, the evidence lower bound (ELBO) is maximized. By letting
\begin{equation}
    \boldsymbol{\mu}_{\boldsymbol{\theta}}(\mathbf{h}_t,t) = \frac{1}{\sqrt{\alpha_t}} \left(\mathbf{h}_t - \frac{1-\alpha_t}{\sqrt{1-\bar{\alpha}_t}} \boldsymbol{\epsilon}_{\boldsymbol{\theta}}(\mathbf{h}_t,t) \right)
\end{equation}
and fixing $\boldsymbol{\Sigma}_{\boldsymbol{\theta}}(\mathbf{h}_t,t) = \tilde{\beta}_t \mathbf{I}$, where $\tilde{\beta}_t =  \frac{1-\bar{\alpha}_{t-1}}{1-\bar{\alpha}_t}\beta_t$, the ELBO objective is reduced to
\begin{equation}
    \sum_{t=1}^T \gamma_t \mathbb{E}_{\mathbf{h}_0,\boldsymbol{\epsilon}} \left[ \| \boldsymbol{\epsilon} - \boldsymbol{\epsilon}_{\boldsymbol{\theta}}(\sqrt{\bar{\alpha}_t} \mathbf{h}_0 + \sqrt{1 - \bar{\alpha}_t} \boldsymbol{\epsilon}, t) \|_2^2 \right]
    \label{eq:objective}
\end{equation}
where $\boldsymbol{\epsilon}_{\boldsymbol{\theta}}(\mathbf{h}_t,t)$ is a time-dependent noise predictor, and $\gamma_{1:T}$ are weight coefficients that depend on $\bar{\alpha}_{1:T}$. Empirically, it is common to set all $\gamma_t = 1$ for better sample quality \cite{ho2020denoising}.

Once the denoising network $\boldsymbol{\epsilon}_{\boldsymbol{\theta}}(\mathbf{h}_t,t)$ is trained, the data generation process begins by sampling $\mathbf{h}_T \sim \mathcal{N}(0,\mathbf{I})$ and then iteratively applies the reverse process for $t = T, \ldots,1$:
\begin{equation}
    \mathbf{h}_{t-1} = \frac{1}{\sqrt{\alpha_t}} \left( \mathbf{h}_t - \frac{1 - \alpha_t}{\sqrt{1 - \bar{\alpha}_t}} \boldsymbol{\epsilon}_{\boldsymbol{\theta}}(\mathbf{h}_t, t) \right) + \sqrt{\tilde{\beta}_t}\mathbf{z}_t,
    \label{eq:ddpm_sampling}
\end{equation}
where $\mathbf{z}_t \sim \mathcal{N}(0,\mathbf{I})$.

\section{Posterior Sampling-Based Channel Estimator}

In this paper, we aim to address the FAS channel estimation problem within the Bayesian framework by leveraging a DM as a learned implicit prior. The model is initially trained offline and then deployed online during the channel estimation stage. To circumvent the intractable computation of the posterior mean, we opt to perform estimation via posterior sampling, which has been shown to achieve near-optimal performance in terms of estimation error \cite{jalal2021robust}.

\subsection{DM as A Learned Prior}

\begin{figure}[t]
\centering
\begin{overpic}[width=0.92\linewidth]{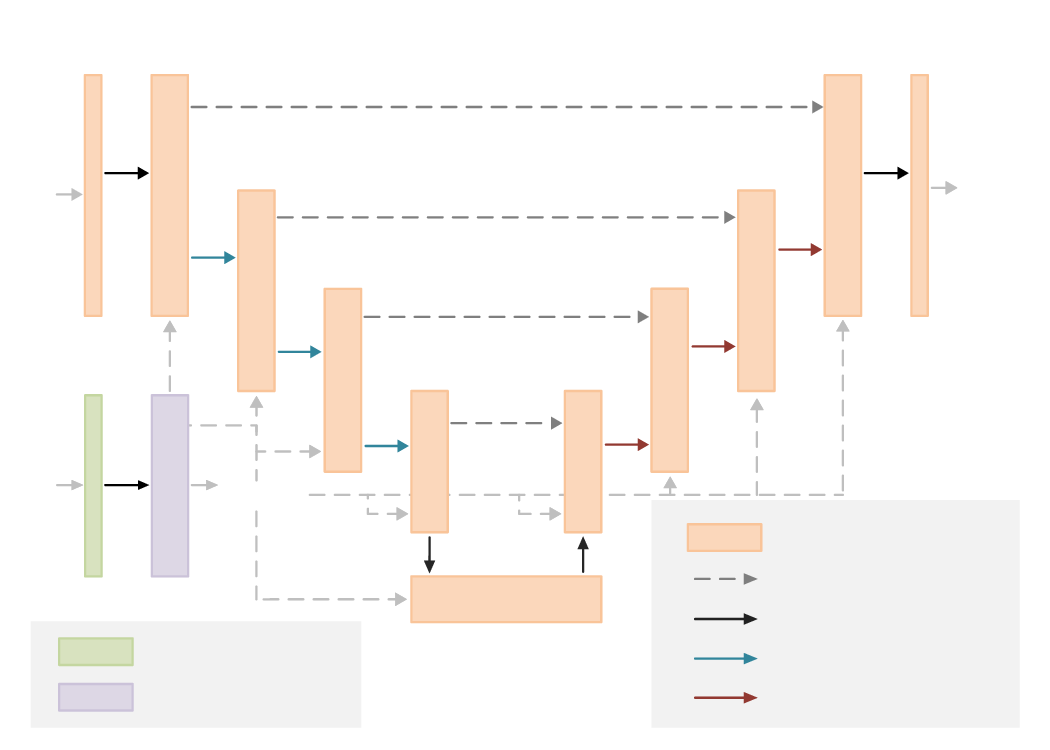}
    \put(3.25,24){\small $t$}
    \put(21.25,24){\small $\mathbf{t}_\mathrm{emb}$}
    \put(1,52){\small $\mathbf{h}_t$}
    \put(92,52.5){\small $\hat{\boldsymbol{\epsilon}}_t$}

    \put(15,8.3){\scriptsize Embedding}
    \put(15,4.1){\scriptsize Linear Layer}

    \put(75,19.6){\scriptsize ResNet block}
    \put(75,15.6){\scriptsize Skip connection}
    \put(75,11.6){\scriptsize Forward pass}
    \put(75,7.6){\scriptsize Downsampling}
    \put(75,3.6){\scriptsize Upsampling}
\end{overpic}
\caption{An illustration of the denoising network $\boldsymbol{\epsilon}_{\boldsymbol{\theta}}$ used in the DM, implemented using a simplified U-Net structure with sinusoidal time embedding.}
\label{fig:nn_architecture}
\end{figure}
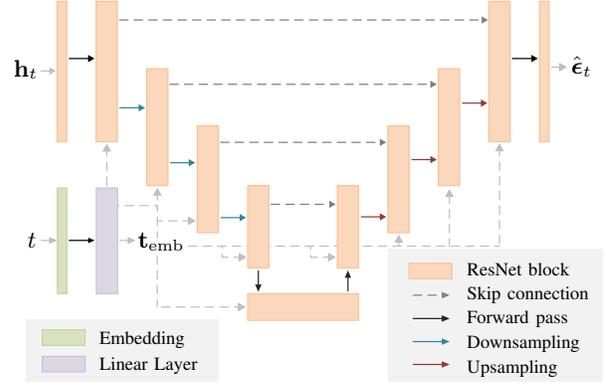

We observe that 2D FAS channel matrices exhibit spatial correlation structures similar to those found in natural images, although characterized by much simpler local patterns. Motivated by this observation, we design a simplified \mbox{U-Net} model for the DM's denoising network $\boldsymbol{\epsilon}_{\boldsymbol{\theta}}$, which accepts the noisy channel $\mathbf{h}_t$ and time step $t$ as inputs and predicts the corresponding noise added $\hat{\boldsymbol{\epsilon}}_t$ as the output. The overall architecture, depicted in Fig.\,\ref{fig:nn_architecture}, adopts an encoder-decoder design with four levels of depth. The encoder progressively downsamples the input using strided convolution while increasing the number of feature channels, whereas the decoder does the opposite via nearest-neighbor interpolation. At each level, two sequential residual (ResNet) blocks are employed for effective feature extraction, and skip connections between the corresponding encoder and decoder levels are used to facilitate the preservation of high-level spatial features.

Since the network operates on tensor-valued variables, the input $\mathbf{h}_t$ is first converted back to the matrix form and its real and imaginary parts are then stacked into two convolutional channels. In addition, the diffusion time step $t$ is encoded via a sinusoidal embedding, followed by a linear projection to produce a time embedding vector $\mathbf{t}_{\mathrm{emb}} \in \mathbb{C}^{D_\mathrm{emb}}$, where $D_\mathrm{emb}$ is the dimension of the embedded space. The embedded time is then passed as another input to each ResNet block to ensure the model is aware of the current diffusion time step.

Given a training dataset $\mathcal{H} = \{\mathbf{h}_0^{(i)}\}_{i=1}^{N_\text{train}}$ with $N_\mathrm{train}$ channel samples, the denoising network is trained offline by minimizing the variational objective defined in~\eqref{eq:objective} using mini-batch stochastic gradient descent (SGD). At each training iteration, a mini-batch of channel realizations $\{\mathbf{h}_0^{(i)}\}_{i=1}^B$ is randomly drawn from the training dataset $\mathcal{H}$, where $B$ denotes the batch size. For each realization $i$, a corresponding noise sample $\boldsymbol{\epsilon}^{(i)}$ and a time-step index $t^{(i)}$ are respectively generated according to $\boldsymbol{\epsilon}^{(i)} \sim \mathcal{N}(\mathbf{0}, \mathbf{I})$ and $t^{(i)} \sim \mathrm{Unif}\{1,\ldots,T\}$. The neural network parameters $\boldsymbol{\theta}$ can then be updated once via gradient descent based on the mini-batch loss
\begin{equation}
    \frac{1}{B} \sum_{i=1}^B \left\lVert \boldsymbol{\epsilon}^{(i)} - \boldsymbol{\epsilon}_{\boldsymbol{\theta}}\left( \sqrt{\bar{\alpha}_{t^{(i)}}} \, \mathbf{h}_0^{(i)} + \sqrt{1 - \bar{\alpha}_{t^{(i)}}} \, \boldsymbol{\epsilon}^{(i)}, t^{(i)} \right) \right\rVert_2^2.
    \label{eq:mini_batch_loss}
\end{equation}
The above training procedure is summarized in Algorithm~\ref{alg:dm_training}.

\addtolength{\topmargin}{0.051in}
\begin{algorithm}[!t]
\caption{Offline DM Training}
\label{alg:dm_training}
\textbf{Require:} Training dataset $\mathcal{H} = \{\mathbf{h}_0^{(i)}\}_{i=1}^{N_{\text{train}}}$ 
\begin{algorithmic}[1]
\Repeat
    \For{$i = 1$ to $B$}
        \State Sample $\mathbf{h}_0^{(i)}$ from $\mathcal{H}$
        \State Sample $t^{(i)} \sim \mathrm{Unif}\{1,\ldots,T\}$
        \State Sample $\boldsymbol{\epsilon}^{(i)} \sim \mathcal{N}(\mathbf{0}, \mathbf{I})$
    \EndFor
    \State Compute the mini-batch loss according to~\eqref{eq:mini_batch_loss}
    \State Update $\boldsymbol{\theta}$ via a gradient descent step
\Until{convergence}
\end{algorithmic}
\end{algorithm}

\subsection{Channel Estimation via Posterior Sampling}

With the trained denoising network $\boldsymbol{\epsilon}_{\boldsymbol{\theta}}(\mathbf{h}_t,t)$ serving as an learned implicit prior for the 2D FAS channel, the channel estimation problem is addressed by sampling from the posterior $p_{\boldsymbol{\theta}}(\mathbf{h}|\mathbf{y})$. To this end, we utilize the framework of the denoising diffusion restoration model (DDRM) \cite{kawar2022denoising}, which defines a conditional generative process through a parametrized conditional Markov chain:
\begin{equation}
    p_{\boldsymbol{\theta}}(\mathbf{h}_{0:T}|\mathbf{y}) = p_{\boldsymbol{\theta}}(\mathbf{h}_T|\mathbf{y}) \prod_{t=1}^T p_{\boldsymbol{\theta}}(\mathbf{h}_{t-1}|\mathbf{h}_t,\mathbf{y}).
    \label{eq:ddrm_reverse}
\end{equation}

The key feature of DDRM is that it defines the forward process in the spectral domain by leveraging the singular value decomposition (SVD) of the measurement matrix. Specifically, for the FAS channel estimation problem, the measurement matrix $\mathbf{S}$ can be decomposed as $\mathbf{S} = \mathbf{I}\boldsymbol{\Sigma}\mathbf{P}$, where $\mathbf{I}$ is an identity matrix, $\boldsymbol{\Sigma}$ is a rectangular diagonal matrix with ones along its main diagonal, and $\mathbf{P}$ is a permutation matrix that reorders the observed ports to the top rows. As a result, the spectral domain simply corresponds to the reordered spatial domain of $\mathbf{h}$. Define the spectral domain representations as $\bar{\mathbf{y}} = \boldsymbol{\Sigma}^\dagger\mathbf{y}$, $\bar{\mathbf{h}} = \mathbf{P}\mathbf{h}$, and $\bar{\mathbf{n}} = \boldsymbol{\Sigma}^\dagger\mathbf{n}$, where $\dagger$ denotes the Moore-Penrose pseudoinverse. Under this transformation, the permuted channel $\bar{\mathbf{h}} \in \mathbb{C}^{\bar{N} \times 1}$ is partitioned as $\bar{\mathbf{h}} = [\mathbf{h}_{\mathcal{O}}^T, \mathbf{h}_{\mathcal{U}}^T]^T$, where $\mathbf{h}_{\mathcal{O}}\in \mathbb{C}^{\bar{M} \times 1}$ and $\mathbf{h}_{\mathcal{U}} \in \mathbb{C}^{(\bar{N}-\bar{M})\times 1}$ correspond to the observed and unobserved ports, respectively. The transformed received signal $\bar{\mathbf{y}} \in \mathbb{C}^{\bar{N} \times 1}$ is constructed by zero-padding $\mathbf{y}$ defined in~\eqref{eq:rx_signal_real} at the unobserved locations, i.e., $\bar{\mathbf{y}} = [\mathbf{y}^T, \mathbf{0}^T]^T$. Similarly, the noise part $\bar{\mathbf{n}} \in \mathbb{C}^{\bar{N} \times 1}$ is expressed as $\bar{\mathbf{n}} = [\mathbf{n}^T, \mathbf{0}^T]^T$. In this form, $\bar{\mathbf{y}}$ contains noisy observations at the observed positions, i.e., $\mathbf{y} = \mathbf{h}_{\mathcal{O}} + \mathbf{n}$, while provides no information about the unobserved ones. The core idea of utilizing DDRM for 2D FAS channel estimation is thus to perform denoising on the elements where observations are available while simultaneously inpainting on the missing ones.

The forward process is constructed as a non-Markovian process conditioned on $\mathbf{y}$:
\begin{equation}
    q(\bar{\mathbf{h}}_{1:T}|\mathbf{h}_0, \mathbf{y}) = q(\bar{\mathbf{h}}_T|\mathbf{h}_0, \mathbf{y}) \prod_{t=1}^T q(\bar{\mathbf{h}}_{t-1}|\mathbf{h}_t, \mathbf{h}_0, \mathbf{y}).
\end{equation}
Let $\bar{\mathbf{h}}_t[i]$ denote the $i$-th element of $\bar{\mathbf{h}}_t$. Each component of the forward process is defined element-wise as follows to reflect the denoising and inpainting strategy described above\footnote{In the interest of brevity, we slightly abuse the notation here by writing $\mathcal{N}(\mu,\sigma)$ for the Gaussian pdf instead of the more precise form $\mathcal{N}(x;\mu,\sigma)$ where the random variable is explicitly indicated.}: \small
\begin{align}
    & q(\bar{\mathbf{h}}_T[i]|\mathbf{h}_0, \mathbf{y}) = \begin{cases}
        \mathcal{N}(\bar{\mathbf{y}}[i],\sigma_T^2 - \sigma_n^2), & \text{if } i < \bar{M} \\
        \mathcal{N}(\bar{\mathbf{h}}_0[i], \sigma_T^2), & \text{Otherwise}
    \end{cases} \\
    & q(\bar{\mathbf{h}}_{t-1}[i]|\mathbf{h}_t, \mathbf{h}_0, \mathbf{y}) = \notag \\
    & \hspace{0.15in} \begin{cases}
        \mathcal{N}(\bar{\mathbf{h}}_0[i] + \sqrt{1-\eta_a^2}\sigma_{t-1}\frac{\bar{\mathbf{y}}[i]-\bar{\mathbf{h}}_0[i]}{\sigma_n}, \eta_a^2 \sigma_{t-1}^2) & \text{if } \sigma_{t-1} \leq \sigma_n \\
        \mathcal{N}((1-\eta_b)\bar{\mathbf{h}}_0[i] + \eta_b \bar{\mathbf{y}}[i], \sigma_{t-1}^2 -\eta_b^2 \sigma_n^2) & \text{if } \sigma_{t-1} \geq \sigma_n \\
        \mathcal{N}(\bar{\mathbf{h}}_0[i] + \sqrt{1-\eta_c^2}\sigma_{t-1}\frac{\bar{\mathbf{h}}_t[i]-\bar{\mathbf{h}}_0[i]}{\sigma_t}, \eta_c^2 \sigma_{t-1}^2) & \text{if } i > \bar{M},
    \end{cases}
\end{align}
\normalsize where $\eta_a, \eta_b, \eta_c \in [0,1]$ are hyperparameters that control the variances and relative weights during the forward diffusion transitions and $\sigma_t^2$ is the variance of the diffusion noise added at each time step. Conceptually, for unobserved elements ($i>\bar{M}$), the noise added is purely contributed by the injected diffusion noise. For the rest of the elements where observations $\mathbf{y}$ are available, the forward diffusion rule further depends on whether the observation noise level $\sigma_n$ exceeds the diffusion noise level $\sigma_{t-1}$. Appropriate weighting and variance adjustments are applied to ensure the consistency of the overall noise level.

The generative process then aims to reverse the forward diffusion process by iteratively predicting the ground truth $\mathbf{h}_0$ from each $\mathbf{h}_t$, which enables an explicit reconstruction of $\bar{\mathbf{h}}_{t-1}$ following the forward process specifications. In particular, the components of the generative process in~\eqref{eq:ddrm_reverse} are defined by simply replacing all ground truth $\bar{\mathbf{h}}_0$ with a predicted one $\bar{\mathbf{h}}_{\boldsymbol{\theta},t}$ for $t<T$, and with $\mathbf{0}$ for $t=T$. This yields the following formulations of the generative process: \small
\begin{align}
    & p_{\boldsymbol{\theta}}(\bar{\mathbf{h}}_T[i]|\mathbf{y}) = \begin{cases}
        \mathcal{N}(\bar{\mathbf{y}}[i],\sigma_T^2 - \sigma_n^2), & \text{if } i < \bar{M} \\
        \mathcal{N}(0, \sigma_T^2), & \text{Otherwise}
    \end{cases} \label{eq:ddrm_initial} \\
    & p_{\boldsymbol{\theta}}(\bar{\mathbf{h}}_{t-1}[i]|\mathbf{h}_t, \mathbf{y}) = \notag \\
    & \hspace{0.05in} \begin{cases}
        \mathcal{N}(\bar{\mathbf{h}}_{\boldsymbol{\theta},t}[i] + \sqrt{1-\eta_a^2}\sigma_{t-1}\frac{\bar{\mathbf{y}}[i]-\bar{\mathbf{h}}_{\boldsymbol{\theta},t}[i]}{\sigma_n}, \eta_a^2 \sigma_{t-1}^2) & \text{if } \sigma_{t-1} \leq \sigma_n \\
        \mathcal{N}((1-\eta_b)\bar{\mathbf{h}}_{\boldsymbol{\theta},t}[i] + \eta_b \bar{\mathbf{y}}[i], \sigma_{t-1}^2 -\eta_b^2 \sigma_n^2) & \text{if } \sigma_{t-1} \geq \sigma_n \\
        \mathcal{N}(\bar{\mathbf{h}}_{\boldsymbol{\theta},t}[i] + \sqrt{1-\eta_c^2}\sigma_{t-1}\frac{\bar{\mathbf{h}}_t[i]-\bar{\mathbf{h}}_{\boldsymbol{\theta},t}[i]}{\sigma_t}, \eta_c^2 \sigma_{t-1}^2) & \text{if } i > \bar{M}, \label{eq:ddrm_general}
    \end{cases}
\end{align}
\normalsize and $\bar{\mathbf{h}}_t$ is mapped back to the original domain via $\mathbf{h}_t = \mathbf{P}^T\bar{\mathbf{h}}_t$.

As demonstrated in \cite{kawar2022denoising}, the meticulous design of the weighting terms and the added noise variances in the forward process leads to a conditional marginal distribution of each latent variable that follows $q(\mathbf{h}_t|\mathbf{h}_0) = \mathcal{N}(\mathbf{h}_t;\mathbf{x}_0|\sigma^2_t\mathbf{I})$, which corresponds to the variance-exploding (VE) formulation of the diffusion process used in an unconditional DM\footnote{On the contrary, the diffusion process we introduced in~\eqref{eq:ddpm_diffusion} is variance-preserving (VP). Both formulations are actually equivalent upon a time-dependent scaling on $\mathbf{h}_t$: $\mathbf{h}_t^{\text{VE}} = \mathbf{h}_t^{\text{VP}}/\sqrt{\bar{\alpha}_t}$.}. Furthermore, under specific choices of the hyperparameters $\{\eta_a,\eta_b, \eta_c\}$, the ELBO objective for unconditional DMs, as defined in~\eqref{eq:objective}, can serve as a valid surrogate for training the conditional DDRM. This implies that a denoising network $\boldsymbol{\epsilon}_{\boldsymbol{\theta}}$ can be trained offline following the unsupervised training procedure outlined in Algorithm~\ref{alg:dm_training}, except that additional transformations to the VE representation and to the spectral domain are needed when applied to the conditional generative process for posterior sampling. In practice, the hyperparameters $\{\eta_a,\eta_b, \eta_c\}$ can be finely tuned to adjust the relative contributions from the observation and generated parts during the sampling process.

A notable feature of the conditional generative process in~\eqref{eq:ddrm_general} is that it explicitly constructs the next latent sample $\bar{\mathbf{h}}_{t-1}$ by first predicting the ground truth sample $\hat{\mathbf{h}}_{\boldsymbol{\theta},t}$ at each time step $t$. This enables an acceleration scheme via skipped sampling, analogous to the denoising diffusion implicit model (DDIM) \cite{song2021denoising} used for unconditional generation. Specifically, given a sub-sequence $\{\tau_i\}_{i=1}^{T'}$ of $\{1,\ldots,T\}$ with $T' \ll T$, which is referred to as the sampling trajectory, the sampling procedure can be sped up by updating only partial latent variables $\{\mathbf{h}_{\tau_1}, \ldots, \mathbf{h}_{\tau_{T'}}\}$. Each update follows~\eqref{eq:ddrm_general} by first predicting the ground truth as
\begin{equation}
    \hat{\mathbf{h}}_{\boldsymbol{\theta}}(\mathbf{h}_{\tau_i}, \tau_i) = \frac{1}{\sqrt{\bar{\alpha}_{\tau_i}}} (\mathbf{h}_{\tau_i} - \sqrt{1-\bar{\alpha}_{\tau_i}}\boldsymbol{\epsilon}_{\boldsymbol{\theta}}(\mathbf{h}_{\tau_i}, \tau_i)),
    \label{eq:gt_prediction}
\end{equation}
followed by a transformation to the VE representation and the spectral domain considered in the posterior sampling process:
\begin{equation}
   \bar{\mathbf{h}}_{\boldsymbol{\theta}, \tau_i} = \mathbf{P}\hat{\mathbf{h}}_{\boldsymbol{\theta}}(\mathbf{h}_{\tau_i},\tau_i)/\sqrt{\bar{\alpha}_{\tau_{i}}}.
\end{equation}
Conversely, a reverse transformation by $ \mathbf{h}_{\tau_{i}} = \sqrt{\bar{\alpha}_{\tau_{i}}}\mathbf{P}^T\bar{\mathbf{h}}_{\tau_{i}}$ is necessary before passed as an input to the pre-trained denoising network $\boldsymbol{\epsilon}_{\boldsymbol{\theta}}$.

The overall algorithm of the posterior sampling procedure for channel estimation is summarized in Algorithm~\ref{alg:dm_sampling}. Note that we let $\tau_0 = 0$ to maintain the consistency of the sampling iteration.

\addtolength{\topmargin}{0.051in}
\begin{algorithm}[!t]
\caption{Channel Estimation by Posterior Sampling}
\label{alg:dm_sampling}
\textbf{Require:} $\mathbf{y}$, $\mathbf{S}$, $\sigma_n$, $\{\bar\alpha_t\}_{t=1}^T$, trained denoising network $\boldsymbol{\epsilon}_{\boldsymbol{\theta}}$, sampling trajectory $\{\tau_i\}_{i=1}^{T'}$
\begin{algorithmic}[1]
\State Initialize $\bar{\mathbf{h}}_{\tau_{T'}}$ by~\eqref{eq:ddrm_initial}
\For{$i = T'$ to $0$}
    \State $\mathbf{h}_{\tau_{i}} \leftarrow \sqrt{\bar{\alpha}_{\tau_{i}}}\mathbf{P}^T\bar{\mathbf{h}}_{\tau_{i}}$
    \State Compute $\hat{\mathbf{h}}_{\boldsymbol{\theta}}(\mathbf{h}_{\tau_i}, \tau_i)$ by~\eqref{eq:gt_prediction}
    \State $\bar{\mathbf{h}}_{\boldsymbol{\theta}, \tau_i} \leftarrow \mathbf{P}\hat{\mathbf{h}}_{\boldsymbol{\theta}}(\mathbf{h}_{\tau_i},\tau_i)/\sqrt{\bar{\alpha}_{\tau_{i}}}$
    \State Update $\bar{\mathbf{h}}_{\tau_{i-1}}$ by~\eqref{eq:ddrm_general}
\EndFor
\State $\hat{\mathbf{h}} \leftarrow \mathbf{P}^T\bar{\mathbf{h}}_{0}$
\State \Return $\hat{\mathbf{h}}$
\end{algorithmic}
\end{algorithm}

\section{Numerical Results}

\enlargethispage{-0.041in}

In this section, we present numerical simulation results to evaluate the performance of our proposed posterior sampling channel estimator. We consider a narrowband system operating at frequency $f_c = 3.5\text{GHz}$. The receiver FAS has $M=4$ RF-chains and $N=51\times51$ ports that are uniformly deployed over a square area of size $W=4\lambda \times 4\lambda$, where the carrier wavelength $\lambda = 1/f_c$. The number of propagation paths is set to $N_p=90$ to indicate a rich scattering environment. For each path $i$, the azimuth and elevation AoAs are assumed to be jointly following $p(\theta_i,\phi_i) = \frac{\cos(\phi_i)}{2\pi}$, which corresponds to the situation where all scatterers are uniformly distributed over the half-spherical region in front of the FAS panel. The complex coefficient of each path is assumed to be independent and identically distributed following $g_i \sim \mathcal{CN}(0,\frac{1}{N_p})$. The simulation is conducted on a server equipped with an NVIDIA A40 GPU and an Intel Xeon Gold 6348 CPU.

During the offline training stage, $N_\text{train} = 40000$ channel realizations are generated according to~\eqref{eq:chan_model} to form the training dataset $\mathcal{H} = \{\mathbf{h}_0^{(i)}\}_{i=1}^{N_{\text{train}}}$. The total diffusion time step is set to $T= 500$ and the noise schedule $\{\beta_t\}_{t=1}^T$ is configured to increase linearly from $\beta_1 = 10^{-4}$ to $\beta_T = 0.02$, which follows the convention established in \cite{ho2020denoising}. The denoising network $\boldsymbol{\epsilon}_{\boldsymbol{\theta}}$ is then trained using Algorithm~\ref{alg:dm_training} with batch size $B=64$, $500$ epochs, and the Adam optimizer with a learning rate of $10^{-4}$.

For online channel estimation, we set $\eta_a=0$, $\eta_b=1$, and $\eta_c=1$ to maximize the contribution from the observation $\mathbf{y}$. In addition, we ignore the variance terms in~\eqref{eq:ddrm_general} to minimize stochasticity during the sampling process, which corresponds to the deterministic sampling introduced in \cite{song2021denoising}.

To verify the effectiveness of our proposed approach, we compare it against the following CS-based baselines:
\begin{itemize}
    \item \textbf{OMP}: The classical Orthogonal Matching Pursuit (OMP) algorithm is applied by exploiting the sparse representation of $\mathbf{h}$ via a 2D discrete Fourier transform (DFT) and assuming the knowledge of $N_p$. The 1D variant of this baseline is discussed in \cite{zhang2024successive}.
    \item \textbf{SBL}: The sparse Bayesian learning approach in \cite{xu2024sparse} aims to find a sparse representation of $\mathbf{h}$ by building an over-complete dictionary in the virtual angle domain and assigns a parametrized Gaussian prior to it. The parameters are iteratively learned via the expectation-maximization (EM) algorithm. We generalize the 1D version of SBL in \cite{xu2024sparse} by considering $G=50$ angular grids for both azimuth and elevation AoAs.
\end{itemize}
We also compare the performance between the accelerated and unaccelerated versions of our posterior sampling scheme. In the accelerated case, only $T'=25$ latent variables are samples and the sampling trajectory $\{\tau_i\}_{i=1}^{T'}$ is constructed by uniformly quantizing $\{1,\ldots,T\}$.

\enlargethispage{-0.041in}

\begin{figure}[!t]
    \centering
    \includegraphics[width=0.92\linewidth]{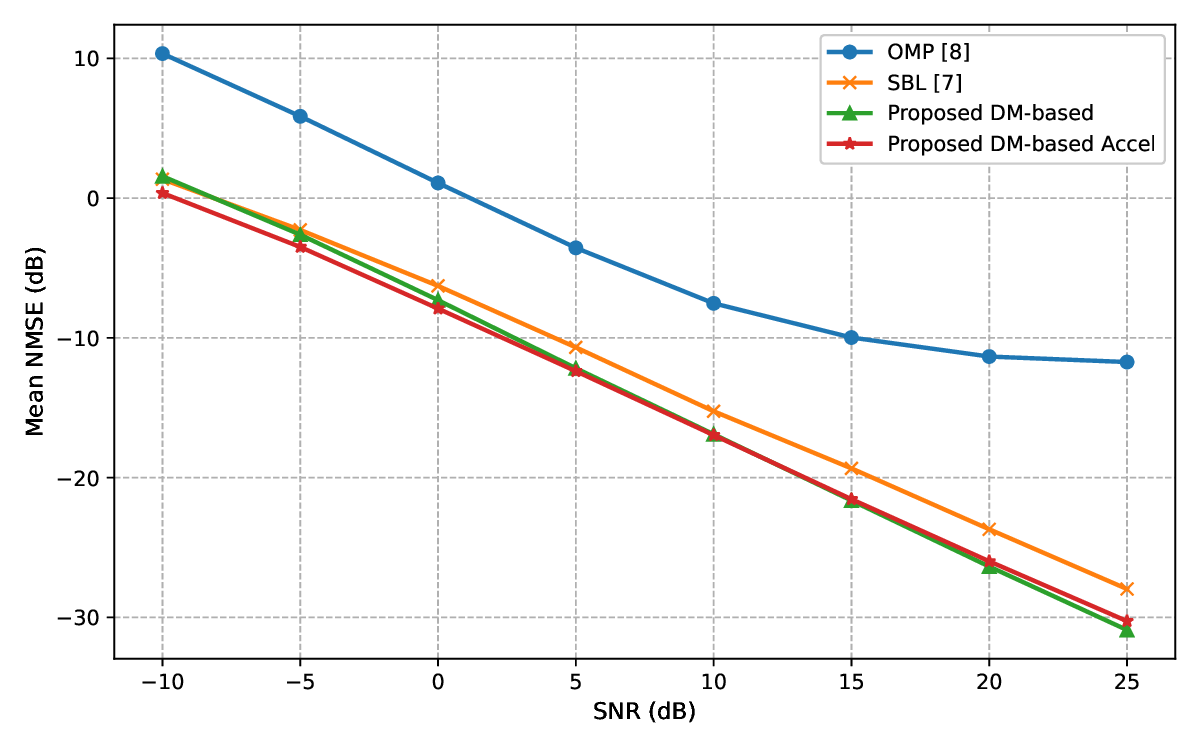}
    \caption{NMSE performance versus SNR.}
    \label{fig:nmse_vs_snr}
\end{figure}

In Fig.\,\ref{fig:nmse_vs_snr}, we compare the normalized mean squared error (NMSE) versus SNR for all methods, with $L=125$ such that the total number of observations is $LM = 500$. The NMSE is defined as
\begin{equation}
    \mathrm{NMSE} = \mathbb{E} \left[ \frac{\|\hat{\mathbf{h}}-\mathbf{h}\|_2^2}{\|\mathbf{h}\|_2^2} \right],
\end{equation}
and the SNR is given by $\mathrm{SNR} = 1/{\sigma_n^2}$ since the channel $\mathbf{h}$ is already normalized according to the construction in~\eqref{eq:chan_model}. As seen in Fig.\,\ref{fig:nmse_vs_snr}, our proposed DM-aided posterior sampling approach outperforms the state-of-the-art SBL method across the entire SNR range. In particular, a wider margin of improvement is achieved at high SNR. We also note that the accelerated scheme exhibits negligible performance loss compared to the unaccelerated one and even achieves a slight improvement in the low SNR region.

\begin{figure}[!t]
    \centering
    \includegraphics[width=0.92\linewidth]{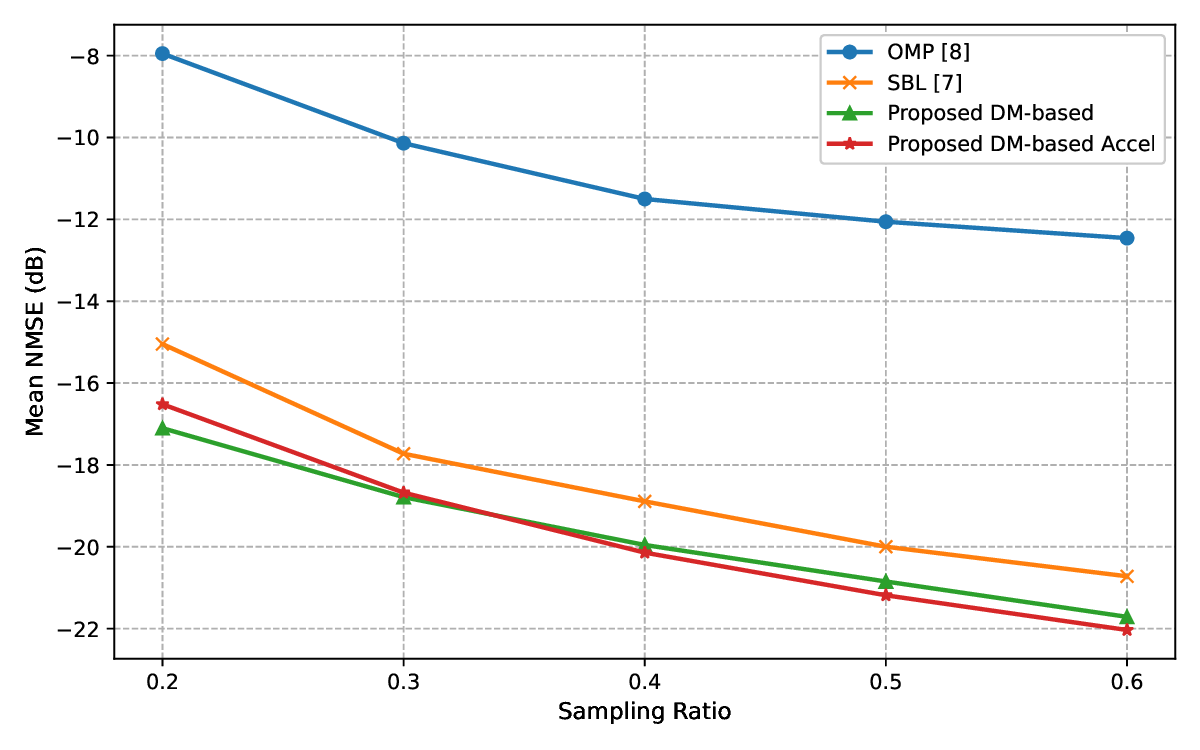}
    \caption{NMSE performance versus sampling ratio.}
    \label{fig:nmse_vs_pilot}
\end{figure}

Fig.\,\ref{fig:nmse_vs_pilot} then shows the NMSE versus sampling ratio performance, fixing the SNR at $10\text{dB}$. The sampling ratio is defined as $\delta = \frac{LM}{N}$. Again, both the accelerated and unaccelerated versions of our proposed method achieve a significant performance gain over the SBL algorithm across the entire range of sampling ratios. An even wider margin can be observed when comparing with the classical OMP algorithm.

\begin{figure}[!t]
    \centering
    \includegraphics[width=0.92\linewidth]{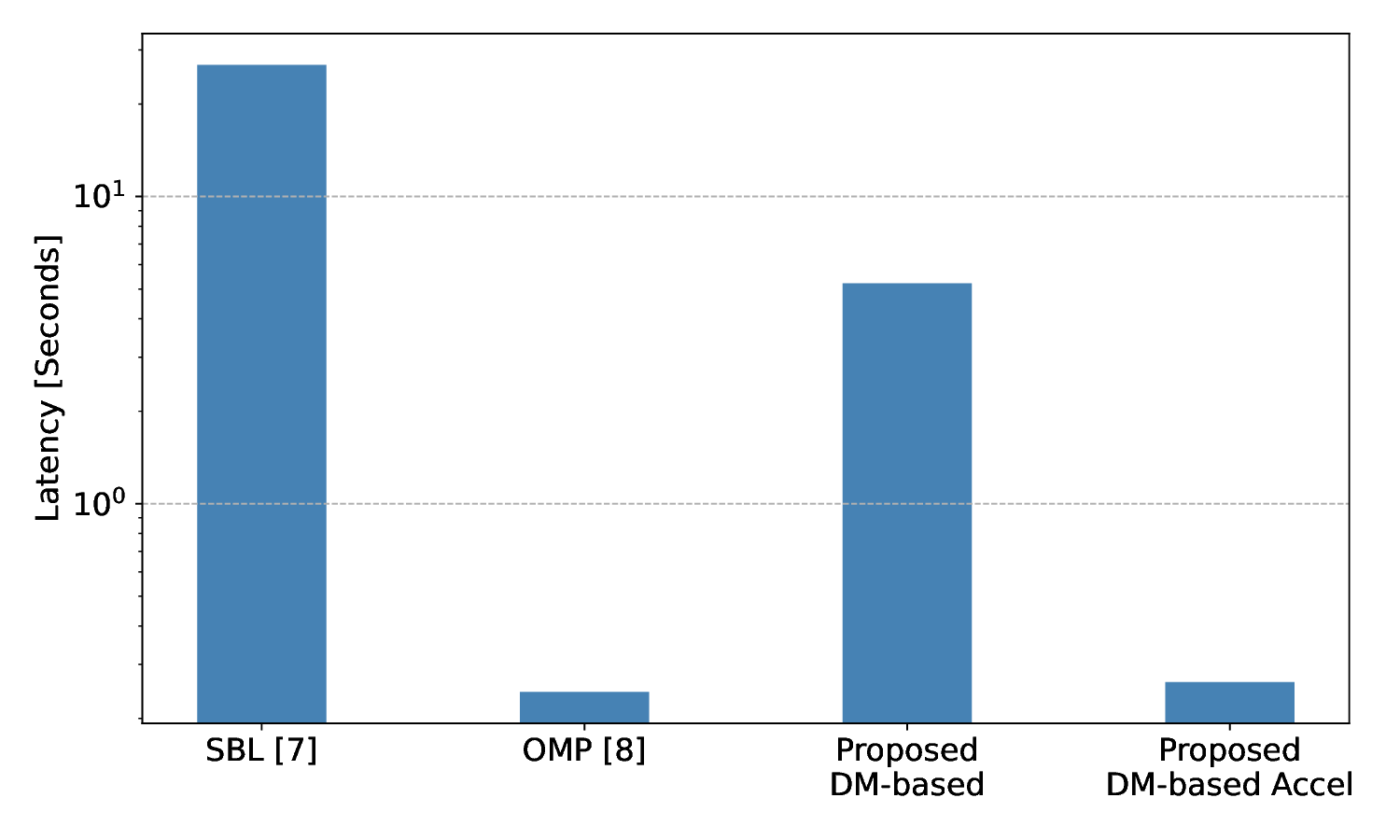}
    \caption{NMSE performance versus sampling ratio.}
    \label{fig:latency}
\end{figure}

Finally, Fig.\,\ref{fig:latency} compares the online channel estimation latency of each method. The figure demonstrates that the state-of-the-art SBL algorithm suffers from prohibitive high latency for 2D FAS channels, which is attributed to the high-dimensional matrix inversion involved in the EM algorithm. On the contrary, our accelerated posterior sampling method achieves a significantly lower computational time, comparable to that of OMP. This highlights the efficiency of our proposed method, which maintains superb and robust estimation performance while significantly reducing the computational complexity.

\section{Conclusion}

\enlargethispage{-0.031in}

In this paper, we proposed an efficient DM-aided channel estimation method for 2D FAS. By leveraging the spatial correlation inherent in FAS channels, our method utilizes a DM trained offline as a learned implicit prior for the channel vector. The DM is subsequently applied online for channel estimation via posterior sampling, during which an acceleration can be achieved by skipped sampling. The accelerated version significantly reduces the computational time while preserving the estimation accuracy, which demonstrates its potential for real-time higher-dimensional FAS channel estimation.

\bibliographystyle{IEEEtran}
\bibliography{main}

\end{document}